\documentclass[preprint,12pt]{aastex}           
\usepackage{natbib,graphics}

\newcommand{\al}{\alpha}
\newcommand{\be}{\begin{equation}}
\newcommand{\ee}{\end{equation}}
\newcommand{\bdm}{\begin{displaymath}}
\newcommand{\edm}{\end{displaymath}}
\newcommand{\bea}{\begin{eqnarray}}
\newcommand{\eea}{\end{eqnarray}}

\newcommand{\Om}{\Omega}
\newcommand{\Omt}{\tilde{\Omega}}
\newcommand{\Jd}{\dot{J}_a}
\newcommand{\Md}{\dot{M}}

\begin{document}

\title{Gravitational Radiation Evolution of Accreting Neutron Stars} 

\author{Robert V. Wagoner\altaffilmark{1}} 
\affil{Dept. of Physics and Center for Space Science and Astrophysics \\ 
Stanford University, Stanford, CA 94305--4060}
\author{Joseph F. Hennawi\altaffilmark{2}}
\affil{Dept. of Astrophysical Sciences, Princeton University, Princeton, NJ 08544--1001}
\and\author{Jingsong Liu\altaffilmark{3}}
\affil{Dept. of Physics, Stanford University, Stanford, CA 94305--4060}

\altaffiltext{1}{wagoner@stanford.edu} 
\altaffiltext{2}{jhennawi@astro.princeton.edu} 
\altaffiltext{3}{jsliu@stanford.edu}

\slugcomment{To appear in the Proceedings of the 20th Texas Symposium on Relativistic Astrophysics}

\begin{abstract}
The gravitational-wave and accretion driven evolution of neutron stars in low mass X-ray binaries and similar systems is analyzed, while the amplitude of the radiating perturbation (here assumed to be an r-mode) remains small. If most of the star is superfluid, with (temperature independent) mutual friction dominating the ordinary (temperature dependent) shear viscosity, the amplitude of the mode and the angular velocity of the star oscillate about their equilibrium values with a period of at least a few hundred years. The resulting oscillation of the neutron star temperature is also computed. For temperature dependent viscosity, the general conditions for the equilibrium to be stable are found.
\end{abstract}


\newpage

\section{Introduction}

We shall study the evolution of rapidly rotating accreting neutron stars under the influence of their emission of gravitational radiation. We modify and extend the two-component model of the star (equilibrium plus perturbation) introduced by \citet{owen98} and also employed by \citet{lev99}, but restrict the analysis to small perturbations. These are assumed to be in the form of r-modes \citep{and98,fm98,lom98,aks99}, which radiate mainly via Coriolis-driven velocity perturbations rather than the density perturbations of the less powerful f-modes. We must allow for large uncertainties in many of the relevant properties of neutron stars, such as the superfluid transition temperature and the effects of magnetic fields and a core-crust boundary layer. 

After developing a general formalism, we shall study the evolution of the accreting neutron star under conditions in which its angular velocity remains approximately constant. One reason for this restriction is our interest in conditions in which the gravitational radiation is persistent over long time scales. This requires the existence of a stable or (possibly) overstable equilibrium. We shall see under what conditions the system can evolve toward this equilibrium, in which the rate of accretion of angular momentum from the surrounding disk is balanced by its rate of loss via gravitational radiation. If this equilibrium is achieved, the observed flux of gravitational radiation can be shown to be proportional to the observed flux of X-rays from the accretion \citep{wag84,bil98}. 

One of our longer term goals is the development of parameterized expressions describing possible time evolutions of the gravitational-wave frequency and amplitude, to facilitate detection by LIGO, VIRGO, and similar laser interferometer detectors. The brightest low mass X-ray binaries (LMXBs) are the prime targets.

\section{Dynamical and Thermal Evolution}

Consider a Newtonian neutron star in equilibrium (with equatorial radius $R$) which is perturbed by a nonaxisymmetric infinitesimal fluid displacement $\vec{\xi}=\vec{f}(r,\theta)e^{i(m\phi +\sigma t)}\sim \alpha R$, with $\al\ll 1$.
Based on the work of \citet{fs78a} and \citet{lu00}, the total angular momentum $J$ of the star can be decomposed into its equilibrium angular momentum $J_*$ and a perturbation proportional to the canonical angular momentum $J_c$. That is,
\be
J = J_*(M,\Om) + (1-K_j)J_c \; , \qquad J_c = -K_c\al^2 J_* \; , \label{decomp}
\ee
where $M$ is the mass and $\Om$ is the (uniform) angular velocity of the equilibrium star. All constants $K_{(\;)}$ will be dimensionless, with $K_j\sim K_c\sim 1$.

In classical mechanics, the action $I=E/\omega$ of any normal mode of a set of oscillators (with frequency $\omega$) is an adiabatic invariant. For a fluid, the analogous quantity should be $\tilde{E}_c/\omega$, where $\tilde{E}_c$ is the canonical energy of the perturbation in the corotating frame and $\omega=\sigma+m\Omega$ is its frequency in that frame. However, we also have the general relation $\tilde{E}_c = -(\omega/m)J_c$ \citep{fs78a}. Therefore, following \citet{hl00}, we assume that the canonical angular momentum is also an adiabatic invariant, and should therefore be unaffected by the slow rate of mass accretion. Thus it obeys the usual relation \citep{fs78b}
\be 
dJ_c/dt = 2J_c[(F_g(M,\Om)-F_v(M,\Om,T_v)] \; ,                  \label{canonical}
\ee
where $F_g$ is the gravitational radiation growth rate and $F_v$ is the viscous damping rate. The latter usually depends upon a spatially averaged temperature $T_v(t)$. 

On the other hand, conservation of total angular momentum requires that
\be
 dJ/dt = 2J_c F_g + \Jd(t) \; ,                               \label{totangmom}
\ee
where $\Jd=j_a\Md$ is the rate of accretion of angular momentum. The mass is accreted with specific angular momentum $j_a$ at a rate $\Md(t)$. 

Combining these equations then gives the dynamical evolution relations
\bea
{1\over\al}{d\al\over dt} & = & F_g-F_v + [K_jF_g+(1-K_j)F_v]K_c\al^2 - \left({j_a\over 2J_*}\right)\Md(t) \; , \label{dadt} \\
\left({I_*\over J_*}\right){d\Om\over dt} & = & -2[K_jF_g+(1-K_j)F_v]K_c\al^2 + \left[{(j_a-j_*)\over J_*}\right]\Md(t) \; ; \label{dodt}
\eea
where $I_*(M,\Om)=\partial J_*/\partial\Om$ and $j_*(M,\Om)=\partial J_*/\partial M$.

In obtaining our thermal evolution relation, the large uncertainties in some thermodynamic properties of the neutron star make it sufficient to consider slowly rotating stars. Thermal energy conservation for the entire star then gives
\be
\int{\partial T\over\partial t}c_v dV \equiv C(T){dT\over dt} \cong 2\tilde{E}_c F_v(T_v) + K_n\langle\Md\rangle c^2 - L_\nu(T_\nu) \; , \label{entire}
\ee
where the rotating frame canonical energy ${\tilde E}_c = K_e\Om J_*\al^2$. 
Since the main contributor to the specific heat is the degenerate relativistic electrons, its value at constant volume ($c_v$) is essentially the same as that at constant pressure. The emissivities are produced by viscous heating, pycnonuclear reactions and neutron emissions in the inner crust (proportional to a time-averaged mass accretion rate), and neutrino losses. The hydrogen/helium burning rate is assumed to be balanced by the surface emission of photons \citep{sch99}, especially at the large accretion rate $\Md=10^{-8}M_\odot\mbox{ yr}^{-1}$ that we shall use. 
The mass accretion rate can be estimated from accretion energy conservation. The photon luminosity arising directly from the accretion is $L_{acc} \approx (GM/R)\Md(t)$, for a slowly rotating neutron star with a negligible magnetosphere.

We are interested in the evolution of neutron stars after they have been spun up to the point where the gravitational radiation growth rate has become equal to the viscous damping rate: 
\be
F_g(\Om_0,M_0)=F_v(\Om_0,M_0,T_0)\equiv F_0 \; ,        \label{initial}
\ee
so the evolution equation (\ref{canonical}) vanishes. This equality defines our initial state. Before that time, we see from equation(\ref{canonical}) that any intrinsic perturbation could not grow from its (infinitesimal) value $\al_{min}$.
The initial temperature $T_0$ is determined by the vanishing of equation (\ref{entire}), with the nuclear heating in the inner crust balanced by the neutrino emission \citep{br00}. 
Since we are only considering conditions in which $\al^2\ll 1$, the properties $\Om(t)$ and $M(t)$ evolve much slower than $\al(t)$ and $T(t)$. For any other property $Q_*$ of the unperturbed star, let $Q_0\equiv Q_*(M_0,\Om_0)$. 

From now on we shall take the perturbation to be due to the dominant $l=m=2$ r-mode. In order to facilitate comparison with previous results, we shall adopt the neutron star model of \citet{owen98} for numerical work. Then the gravitational radiation growth rate of this mode is 
\be 
F_g = \Omt^6/\tau_{gr} \; , \quad \tau_{gr} = 3.26 \mbox{ sec} \; , 
\qquad \Omt\equiv\Om(\pi G\langle\rho\rangle)^{-1/2} \; . \label{Fg}
\ee
In the temperature range of interest ($10^8 < T < 10^{10}$ K), the viscous damping rate of this mode is approximated as
\be
F_v \cong \left(\frac{\Omt^5}{\tau_{mf}}\right)e^{-(T_v/T_c)^2} + \frac{1}{\tau_{sh}}\left(\frac{10^9\mbox{ K}}{T_v}\right)^2 \; , \label{Fv}
\ee
where $T_c$ is the superfluid transition temperature.

The first term is the contribution from the mutual friction between the neutron superfluid and the superconducting proton--relativistic electron fluid \citep{lm00}. Its behavior (and that of other properties considered below) as the temperature passes through the superfluid transition temperature $T_c$ is approximated by the exponential. Although \citet{lm00} found that $\tau_{mf}\la 10^4$ sec, we shall keep it as a free parameter because of the many uncertainties involved in the physics of this system, especially when including magnetic effects such as the interaction of the vortex lines and flux tubes \citep{rud}. In fact, we fix it to satisfy our initial condition (\ref{initial}), when observationally relevant values of $\Om_0,M_0,T_0$ are chosen. 
To approximately match an inferrred maximum spin rate of 330 Hz for the neutron stars in LMXBs, we shall choose $\Omt_0 = \Om_0(\pi G\langle\rho\rangle)^{-1/2} = 0.25$ \citep{lev99}, which then fixes $\tau_{mf}\approx 13$ sec if $T\ll T_c$. The maximum rotation rate of a neutron star corresponds to $\Omt\cong 2/3$.

The second term is the contribution of the ordinary shear viscosity. \citet{lm00} obtained $\tau_{sh} = 1.0\times 10^8$ sec for their superfluid neutron star model, dominated by electron--electron scattering in the superfluid regions and neutron--neutron scattering in the normal regions. Above the superfluid transition temperature, \citet{lom98} obtained $\tau_{sh} = 2.5\times 10^8$ sec. There are also large uncertainties in this contribution, due to the shear in a boundary layer between the core and crust \citep{ajks,bu00,lou00,rie,wu} and the uncertain response of the crust to the mode \citep{lu01}.

In keeping with the fact that we are only working to lowest order in $\Omt$ (as well as the relativity parameter $GM/Rc^2$), we take $j_a-j_0\cong j_a$ and $J_0\cong I_0\Om_0$. We also note that $K_c = 3K_e = 0.094$. (The value of $K_j$ is unimportant.) 

Now that we have specified all properties in the equations (\ref{dadt}) and (\ref{dodt}) of evolution of $\al(t)$ and $\Om(t)$, we can consider the thermal evolution. In what follows we shall assume that thermal conductivity timescales are short enough to give relations $T_v(T)$ and $T_\nu(T)$ between these three spatially averaged temperatures that appear in equation (\ref{entire}). The normal and superfluid contribution to the specific heat and the neutrino luminosity that appear in this equation are approximated by 
\bea
C(T) & = & [C'_{norm}e^{-(T_c/T)^2} + C'_{super}]T \qquad (C'_{norm} \cong 20C'_{super}) \; ,\\
L_\nu(T) & = & L'_{URCA}T^8e^{-(T_c/T)^2} + L'_{brem}T^6 \; .   
\eea
The constants $L'_{URCA}$ and $L'_{brem}$ are obtained by fitting the results of \citet{br00} for normal and superfluid neutron stars. We also take the nuclear heating constant $K_n = 1\times 10^{-3}$ \citep{br00}.

In Figures 1 and 2 we show the results of integrating our three coupled evolution equations (\ref{dadt}), (\ref{dodt}), and (\ref{entire}), if $T_c\gg T_0$. The main feature is the spin-down (due to loss of angular momentum in gravitational waves) and viscous heating during the time when the amplitude $\al$ of the mode exceeds its equilibrium value (denoted by the dashed line). 

\begin{figure}[!ht]
\plottwo{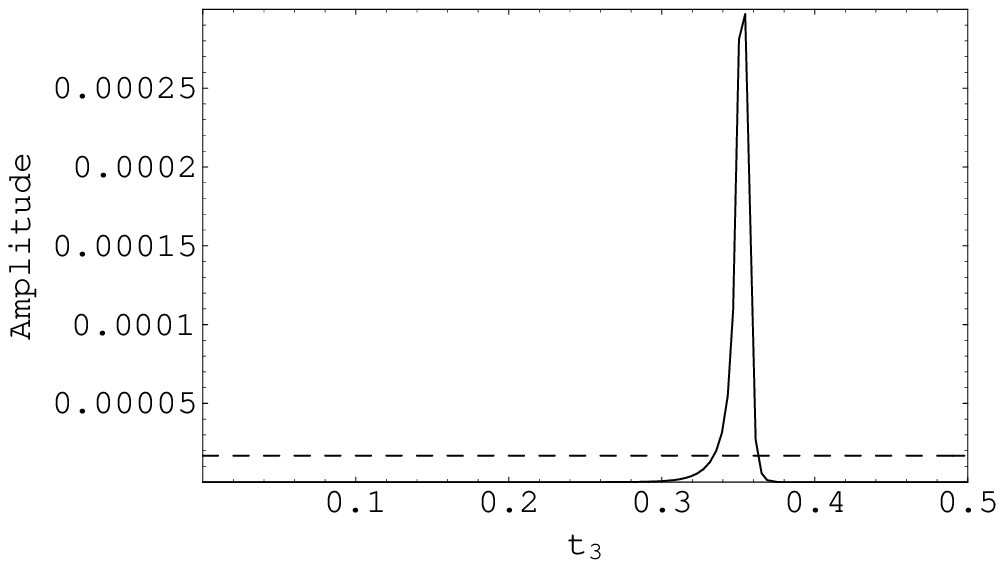}{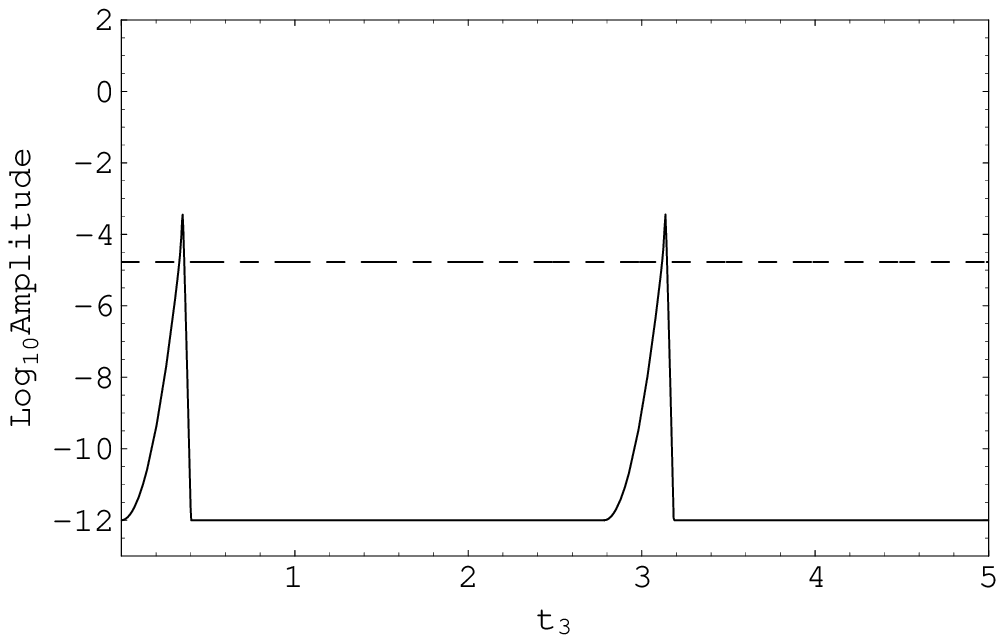}
\caption{The evolution of the mode amplitude $\alpha$, for the case $T_c\gg T_0$. The dashed line is its equilibrium value. Time is in units of $10^3$ years. Subsequent cycles are similar.}
\end{figure}
\begin{figure}[!ht]
\plottwo{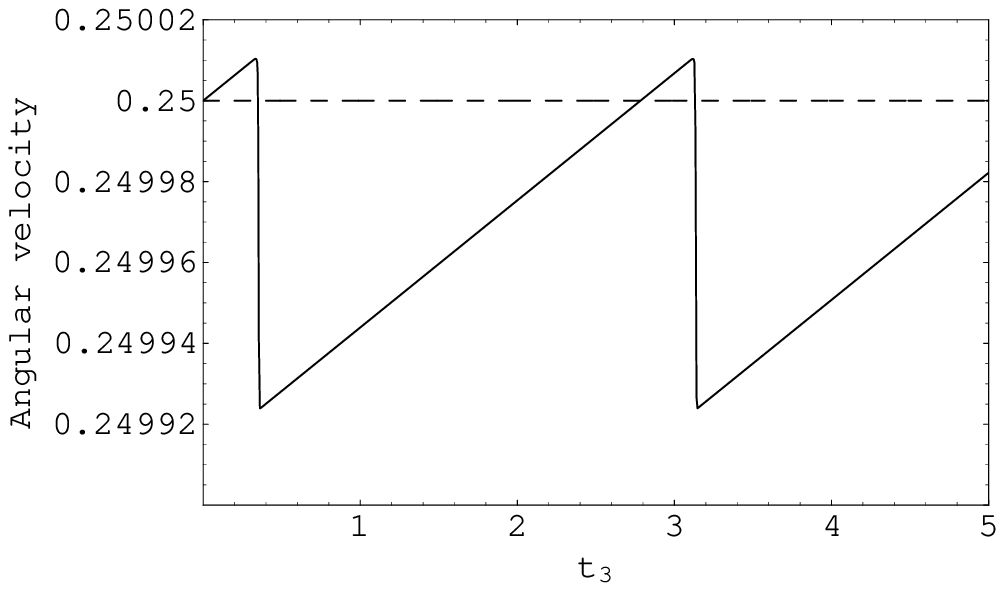}{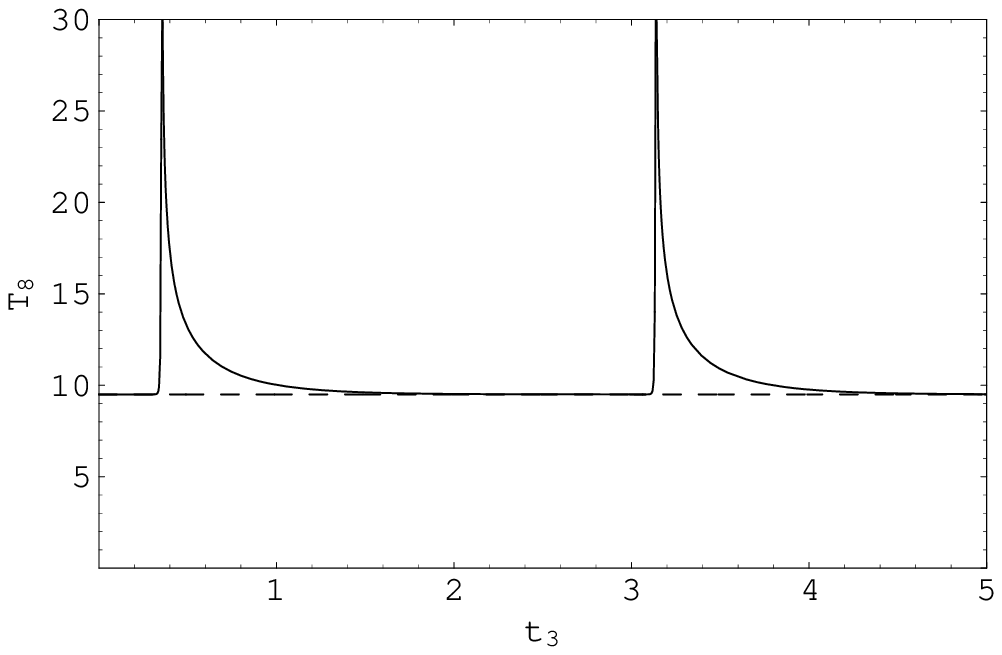}
\caption{Same as Figure 1, for the angular velocity $\Om(\pi G\langle\rho\rangle)^{-1/2}$ and the temperature $T/10^8$ K. Here the dashed line is the initial value.} 
\end{figure}

For such large values of $T_c$, the evolution qualitatively resembles that which occurs when $\partial F_v/\partial T = 0$ (requiring neglect of the shear viscosity). In this case equations (\ref{dadt}) and (\ref{dodt}) decouple from equation (\ref{entire}). In addition, $x\equiv\ln\al$ completely decouples, obeying the equation 
\be
d^2x/dt^2 - 2K_cF_0e^{2x}dx/dt + {dV/dx} = 0\; , \quad V(x)\cong F_0(K_cF_0e^{2x} - F_ax) \; , 
\ee
where $F_a$ is the time-averaged rate of accretion of angular momentum. The Eddington limit to the accretion rate gives $\tau_a\equiv 1/F_a \ga 5\times 10^6$ years. 
The sign of the damping term is opposite to that obtained by \citet{lev99}, leading to overstable oscillations about the equilibrium amplitude 
\bdm
\al_e\cong (F_a/2K_cF_0)^{1/2} \la 2\times 10^{-5}
\edm 
with a period 
\bdm
P\cong [8\ln(\al_{max}/\al_{min})/(F_0F_a)]^{1/2} \ga 300 \mbox{ years} \; .
\edm
The fraction $F_>$ of the time that $\al > \al_e$ is 
\bdm 
F_> \cong \ln[8\ln(\al_{max}/\al_{min})]/4\ln(\al_{max}/\al_{min})\sim 0.1 \; .
\edm
It is also found that $\al_{max}$ increases and $\al_{min}$ decreases on a timescale $\sim P/\al_{max}$. However, $\al_{min}$ is presumably limited by the intrinsic fluctuations in the neutron star, which is also why we have not allowed $\al$ to drop below its initial value in Figure 1. 
Because of the inclusion of the shear viscosity, the value of the period $P$ in Figure 1 is about ten times longer than the above estimate while the value of the fraction $F_>$ in Figure 1 is about ten times less.  

On the other hand, for values of $T_c\la T_0$, the temperature dependence of the viscous damping rate $F_v$ is strong enough to produce a thermal runaway to large values of $T$ and $\al$ (outside the range of validity of our equations), as found by \citet{lev99}. This result is generalized in the next section.

\section{Behavior Near Equilibrium}

In contrast to the initial state, the equilibrium state $X^i_e$ of our dynamical variables 
\bdm
X^i(t)=\{\al,\Om,T\}=X^i_e[1+\zeta^i(t)]\; ,\qquad |\zeta^i|\ll 1 \; ,
\edm
is defined by the vanishing of the evolution equation (\ref{totangmom}), in addition to equations (\ref{canonical}) and (\ref{entire}). 
Employing a constant (averaged) accretion rate, the evolution equations give 
\be
d\zeta^i/dt=A^{ij}\zeta^j\; , \qquad \zeta^i\propto\exp(\lambda t)\; , \quad 
||A^{ij}-\lambda\delta^{ij}|| = 0 \; . 
\ee
Assume now that $|\partial F_v/\partial T|\sim F_v/T_e$, etc. 

The coefficients of the eigenvalue equation $\lambda^3+a_2\lambda^2+a_1\lambda+a_0=0$ are
\bea
a_2 & \cong & {1\over C_e}\left[\left({\partial L\over\partial T}\right)_e - 2(\tilde{E}_c)_e\left({\partial F_v\over\partial T}\right)_e\right]\sim K_r\al_e^2F_0 \; , \\
a_1 & \cong & {4(\tilde{E}_c)_eF_0\over C_e}\left({\partial F_v\over\partial T}\right)_e \sim K_r\al_e^2F_0^2 \; , \\
a_0 & \cong & {4K_c\Om_e\al_e^2F_0\over C_e}\left[{\partial(F_g-F_v)\over\partial\Om}\right]_e\left({\partial L\over\partial T}\right)_e \nonumber \\
{ } & - & {16K_c(\tilde{E}_c)_e\al_e^2F_0^2\over C_e}\left({\partial F_v\over\partial T}\right)_e \sim K_r\al_e^4F_0^3 \; .
\eea
The ratio of rotational to thermal energy is $K_r \equiv 2K_e\Om_eJ_0/C_eT_e\sim 10^5$. We have used the fact that the cooling rate $F_c \equiv L_\nu(T_e)/C_eT_e \sim K_r\al_e^2F_0$.

Now we also employ the inequalities $K_r\gg 1$ and (mode energy)/(thermal energy) $\sim K_r\al_e^2 \la 10^{-4}\Rightarrow |a_1|\gg a_2^2$  to obtain the eigenvalues
\bdm
\lambda_{1,2}\cong -a_2/2\pm\sqrt{-a_1}\; ,\qquad \lambda_3\cong -a_0/a_1 \; .
\edm
We have used the fact that $|\lambda_3|\sim\al_e^2 F_0\ll |\lambda_1|\sim|\lambda_2|$.

Let us examine the two relevant possibilities. For cases such as dominance by shear viscosity in equation (\ref{Fv}), with $\tilde{E}_c>0$, 
\bdm
a_1\propto (\tilde{E}_c)_e(\partial F_v/\partial T)_e < 0 \;\Longrightarrow\; \lambda_{1,2}\cong \pm\sqrt{-a_1} \sim K_r^{1/2}\al_eF_0 \; .
\edm
Thus this equilibrium is unstable, with a growth rate $\lambda_1$ that is of the same magnitude as found by Levin (1999). 

The other possibility is
\bdm
a_1 > 0 \;\Longrightarrow\; \lambda_{1,2}\cong -a_2/2\pm i\sqrt{a_1} \; .
\edm
Thus stability also requires that $a_0>0$ and $a_2>0$. From their relations above, we see that this means that the variation in the cooling rate with temperature must be greater than twice the variation in the viscous heating rate with temperature (which is usually satisfied). 

\acknowledgments

   This work was supported in part by NSF grant PHY--0070935 to R.V.W. and NASA grant NAS 8-39225 to Gravity Probe B. R.V.W thanks the Aspen Center for Physics for support during a 1999 summer workshop, and benefitted from many discussions during the 2000 program on Spin and Magnetism in Young Neutron Stars at the Institute for Theoretcial Physics, U.C. Santa Barbara.


\begin{thebibliography}{99}

\bibitem[Andersson(1998)]{and98} Andersson, N. 1998, \apj, {\bf502}, 708

\bibitem[Andersson, Kokkotas \& Schutz(1999)]{aks99} Andersson, N., Kokkotas, K.D. \& Schutz, B.F. 1999 \apj, {\bf510}, 846

\bibitem[Andersson, Jones, Kokkotas \& Stergioulas(1999)]{ajks} Andersson, N., Jones, D.I., Kokkotas, K.D. \& Stergioulas, N. 1999, \apjl, {\bf534}, L75

\bibitem[Bildsten(1998)]{bil98} Bildsten, L. 1998, \apjl, {\bf501}, L89

\bibitem[Bildsten \& Ushomirsky(2000)]{bu00} Bildsten, L. \& Ushomirsky, G. 2000, \apjl, {\bf529}, L33

\bibitem[Brown(2000)]{br00} Brown, E.F. 1999, \apj, {\bf531}, 988

\bibitem[Friedman \& Morsink(1998)]{fm98} Friedman, J.L. \& Morsink, S.M. 1998, \apj, {\bf502}, 714

\bibitem[Friedman \& Schutz(1978a)]{fs78a} Friedman, J.L. \& Schutz, B.F. 1978, \apj, {\bf221}, 937

\bibitem[Friedman \& Schutz(1978b)]{fs78b} Friedman, J.L. \& Schutz, B.F. 1978, \apj, {\bf222}, 281

\bibitem[Ho \& Lai(2000)]{hl00} Ho, W.C.G. \& Lai, D. 2000, \apj, {\bf543}, 386

\bibitem[Levin(1999)]{lev99} Levin, Y. 1999, \apj, {\bf517}, 328

\bibitem[Levin \& Ushomirsky(2001a)]{lu00} Levin, Y. \& Ushomirsky, G. 2001a, \mnras, {\bf322}, 515

\bibitem[Levin \& Ushomirsky(2001b)]{lu01} Levin, Y. \& Ushomirsky, G. 2001b, \mnras, submitted (astro-ph/0006028)

\bibitem[Lindblom \& Mendell(2000)]{lm00} Lindblom, L \& Mendell, G. 2000, \prd, {\bf61}, 104003

\bibitem[Lindblom, Owen \& Morsink(1998)]{lom98} Lindblom, L., Owen, B.J. \& Morsink, S.M. 1998, \prl, {\bf80}, 4843

\bibitem[Lindblom, Owen \& Ushomirsky(2000)]{lou00} Lindblom, L., Owen, B.J. \& Ushomirsky, G. 2000, \prd, {\bf62}, 084030 

\bibitem[Owen et al.(1998)]{owen98} Owen, B.J., Lindblom, L., Cutler, C., Schutz, B.F., Vecchio, A. \& Andersson, N. 1998, \prd, {\bf58}, 084020

\bibitem[Rieutord(2000)]{rie} Rieutord, M. 2000, \apj, {\bf550}, 443

\bibitem[Ruderman(2000)]{rud} Ruderman, M. 2000, private communication 

\bibitem[Schatz et al.(1999)]{sch99} Schatz, H., Bildsten, L., Cumming, A. \& Wiescher, M. 1999, \apj, {\bf524}, 1014

\bibitem[Wagoner(1984)]{wag84} Wagoner, R.V. 1984, \apj, {\bf278}, 345

\bibitem[Wu, Matzner \& Arras(2001)]{wu} Wu, Y., Matzner, C.D. \& Arras, P. 2001, \apj, {\bf549}, 1011

\end{thebibliography}
\end{document}